\documentclass[preprint2]{aastex}
\usepackage{rotating}
\usepackage{graphicx}
\usepackage{grffile}

\begin{document}
\setcounter{secnumdepth}{2}     
\title{The TRENDS High-Contrast Imaging Survey. II. \\ Direct Detection of the HD 8375 Tertiary}
\author{Justin R. Crepp\altaffilmark{1}, John Asher Johnson\altaffilmark{2,3}, Andrew W. Howard\altaffilmark{4}, Geoff W. Marcy\altaffilmark{5}, Debra A. Fischer\altaffilmark{6}, Scott M. Yantek\altaffilmark{2}, Jason T. Wright\altaffilmark{7,8}, Howard Isaacson\altaffilmark{5}, Ying Feng\altaffilmark{7,8}}
\altaffiltext{1}{Department of Physics, University of Notre Dame, 225 Nieuwland Science Hall, Notre Dame, IN 46556, USA} 
\altaffiltext{2}{Department of Astronomy, California Institute of Technology, 1200 E. California Blvd., Pasadena, CA 91125, USA} 
\altaffiltext{3}{NASA Exoplanet Science Institute (NExScI), CIT Mail Code 100-22, 779 South Wilson Avenue, Pasadena, CA 91125, USA} 
\altaffiltext{4}{Institute for Astronomy, University of Hawaii, 2680 Woodlawn Drive, Honolulu, HI 96822} 
\altaffiltext{5}{Department of Astronomy, University of California, Berkeley, CA 94720, USA}
\altaffiltext{6}{Department of Physics, Yale University, New Haven, CT 06511, USA} 
\altaffiltext{7}{Department of Astronomy \& Astrophysics, The Pennsylvania State University, University Park, PA 16802, USA} 
\altaffiltext{8}{Center for Exoplanets and Habitable Worlds, The Pennsylvania State University, University Park, PA 16802, USA}
\email{jcrepp@nd.edu} 

\begin{abstract}  
We present the direct imaging detection of a faint tertiary companion to the single-lined spectroscopic binary HD~8375 AB. Initially noticed as an 53 $\rm m\:s^{-1}yr^{-1}$ Doppler acceleration by \citet{bowler_10}, we have obtained high-contrast adaptive optics (AO) observations at Keck using NIRC2 that spatially resolve HD~8375~C from its host(s). Astrometric measurements demonstrate that the companion shares a common proper-motion. We detect orbital motion in a clockwise direction. Multiband relative photometry measurements are consistent with a spectral-type of M1V. Our combined Doppler and imaging observations place a lower-limit of $m \geq0.297M_{\odot}$ on its dynamical mass. We also provide a refined orbit for the inner pair using recent RV measurements obtained with HIRES. HD~8375 is one of many triple-star systems that are apparently missing in the solar neighborhood.
\end{abstract}

\section{INTRODUCTION}\label{sec:intro}
We have recently commenced an interdisciplinary program that combines the radial velocity (RV) method with high-contrast imaging to identify faint companions in wide orbits around nearby stars \citep{crepp_12b}. Using approximately $7-25$ years of precise Doppler measurements, we select targets for follow-up adaptive optics (AO) observations based on the existence of long-term RV accelerations (trends), which indicate the presence of a distant body. The objective of the TRENDS ({\bf T}a{\bf R}getting b{\bf EN}chmark-objects with {\bf D}oppler {\bf S}pectroscopy) high-contrast imaging program is to directly detect and study in detail low-temperature bodies that are found to orbit each target star. 

\begin{table}[!t]
\centerline{
\begin{tabular}{lc}
\hline
\hline
          \multicolumn{2}{c}{HD~8375 System Properties}     \\
\hline
\hline
right ascension [J2000]            &     01 23 37.5         \\
declination [J2000]                   &     +34 14 45.2       \\
$B$                                  &    $7.1$                     \\
$V$                                  &     $6.3$                     \\
$J$                                  &    $4.820\pm0.037$  \\
$H$                                 &   $4.222\pm0.236$ \\
$K_s$                             &   $4.290\pm0.023$  \\
d [pc]                                       &    $56.7\pm1.3$   \\
proper motion [mas/yr]        &     $233.1\pm0.3$ E \\
                                                  &      $117.8\pm0.2$ N  \\
\hline
\hline
\end{tabular}}
\caption{Coordinates, apparent magnitudes, distance, and proper motion of HD~8375 from SIMBAD. Magnitudes are from 2MASS \citep{skrutskie_06}. The parallax-based distance is from {\it Hipparcos} measurements using the refined data reduction of \citealt{van_leeuwen_07}.}
\label{tab:starprops}
\end{table}

Once a companion is identified, its relative position on the sky is measured over multiple epochs with imaging, and the primary star is monitored at increased cadence with continued Doppler measurements. Together, these observations can determine the three-dimensional orbit and precise dynamical mass of the companion \citep{crepp_12a}. Spectro-photometric measurements acquired across a wide bandpass may then be used to test theoretical spectral models and thermal evolutionary models by comparing the inferred mass to that found using Newton's laws \citep{boden_06}. The ultimate goal of the TRENDS program is to determine the first orbit (all 6 elements) and dynamical mass of a directly-imaged gas giant extrasolar planet.

The majority of our targets are nearby ($d\lesssim100$ pc), bright ($V\lesssim12$), main-sequence FGK stars with RV time-baselines of at least several years. We also monitor M-dwarfs \citep{apps_10}, massive subgiants \citep{johnson_11}, and young stars (Hillenbrand et al., in prep.). In many cases, Doppler measurements span more than a decade (e.g., \citealt{wright_09}). A number of systems already show curvature (change in the RV acceleration), making it possible to characterize the orbit rapidly and more accurately than astrometry-only surveys \citep{konopacky_10,dupuy_10}.

The TRENDS survey began observations of an intrinsically companion-rich sample of stars in May 2010 at Keck using NIRC2 \citep{crepp_12b}. Many targets have a large proper-motion and follow-up observations demonstrate that their companions are co-moving. In this paper, we report the direct detection of HD~8375~C, the tertiary companion of a single-lined spectroscopic binary. A Doppler acceleration of $53 \; \rm m \: \rm s^{-1} yr^{-1}$was initially noticed by \citet{bowler_10} over a 5.3 year base-line. We show that HD~8375~C is responsible for the trend. Our recent RV measurements improve the orbit solution for HD~8375~AB and extend the baseline of the observed acceleration.

It is known that the census of triple-stars is currently incomplete beyond 10 pc \citep{tokovinin_04}. At a distance of $56.7\pm1.3$ pc, HD~8375 (Table 1) represents one of many hierarchical systems that are apparently missing in the solar neighborhood, having thus far evaded detection by various faint-companion search techniques. Further, few triple systems have had their orbital dynamics studied in detail \citep{duchene_06,schaefer_12}. And, determining the configuration of multi-star systems has important implications for our understanding of star-formation theory, which is constrained primarily from observations of single and binary stars \citep{crepp_10,bate_09}. Given the benefits of combining two complementary techniques, such as precise Doppler measurements and direct imaging, the HD~8375 system provides an excellent opportunity to characterize a hierarchical triple-star system in exquisite detail. 

\section{OBSERVATIONS}
\subsection{Doppler Measurements}
HD~8375 was initially noticed as an (SB1) RV variable star by \cite{beavers_eitter_86}. Also identified as an evolved star by \citealt{deMedeiros_99}, and \citealt{snowden_05}, HD~8375~AB has a designated spectral type of G8IV. Most recently, it was targeted by \citet{johnson_06} at Lick Observatory as part of a dedicated Doppler survey of subgiants to search for extrasolar planets. From this data set the first accurate orbit for HD~8375~AB was published by \citet{bowler_10}. In addition to a large SB1 signal ($K=4939.2^{+2.6}_{-2.5}$ m/s), a subtle long-term acceleration, $\mbox{dv/dt}=52.9^{+1.8}_{-1.9}$ m$\:$s$^{-1}$yr$^{-1}$, suggested that the HD~8375~AB pair is orbited by a tertiary companion. 

\begin{table}[!t]
\centerline{
\begin{tabular}{lcc}
\hline
\hline
      HJD           &  RV                    &  Uncertainty \\
    -2,450,000   &  [m~s$^{-1}$]   &    [m~s$^{-1}$]     \\
\hline
     3,256.962   &  -3,807.92  &    6.67      \\
     3,256.984   &  -3,778.30   &   6.48  \\
     3,326.717   &  -3,359.07   &   5.39  \\
     3,326.737   &  -3,360.36   &   6.14  \\
     3,327.756   &  -3,551.16   &   6.72  \\
     3,327.778   &  -3,553.56   &   6.91  \\
     3,577.974   &  -3,199.62   &   8.29  \\
     3,577.994   &  -3,196.42   &   7.06  \\
     3,602.865   &  -1,058.33   &   5.53  \\
     3,602.888   &  -1,058.29   &   5.52  \\
     3,641.865   &  3,260.39    &  5.07  \\
     3,641.882   &  3,256.72    &  5.24  \\
     3,669.696   &  -4,161.93   &   8.64  \\
     3,708.725   &  5,528.20    &  6.24  \\
     3,708.750   &  5,530.83    &  5.49  \\
     3,976.821   &  3,586.67    &  5.38  \\
     3,998.874  &  -3,360.45    &  6.62  \\
     4,021.860  &  -1,261.40    &  9.15  \\
     4,034.861  &   3,436.29    &  7.47  \\
     4,070.777   &  78.87        &  6.61  \\
     4,073.672  &   -928.42    &   5.02   \\
     4,092.661  &  -4,071.47   &  10.18  \\
     4,135.638  &   5,562.40    &  4.33  \\
     4,136.638  &  5,448.56    &  4.45  \\
     4,150.645  &   1,581.30    &  5.54  \\
     4,170.617  &  -3,921.77    &  6.86  \\
     4,170.633  &  -3,957.81    &  7.71  \\
     4,274.979  &   -769.69    &  5.91   \\
     4,274.996   &  -770.38    &  7.26  \\
     4,288.965  &   4,186.74   &   5.55  \\
     4,737.842  &  1,824.03   &   6.16  \\
     5,060.954  &   5,510.28   &   6.20   \\
     5,060.968  &   5,501.75   &   6.86  \\
     5,091.868  &  -3,448.10   &   6.65  \\
     5,091.881  &  -3,452.50   &   6.45  \\
     5,109.840  &  -2,046.18   &   7.71  \\
     5,109.873  &  -2,015.68   &   7.99  \\
     5,148.778  &  4,712.08   &   7.27  \\
     5,148.794  &  4,717.75   &   7.42  \\
\hline
\hline
\end{tabular}}
\caption{Lick RV measurements.}
\label{tab:keck_hires}
\end{table}

\begin{table}[!t]
\centerline{
\begin{tabular}{lcc}
\hline
\hline
      HJD           &  RV                    &  Uncertainty \\
    -2,450,000   &  [m~s$^{-1}$]   &    [m~s$^{-1}$]     \\
\hline
     6098.133   &    -4823.01   &  1.37    \\         
     6100.109   &   -5124.58    &  1.37   \\
     6149.026   &    4351.28    &  1.08   \\
     6154.045   &    3643.48    &  1.03  \\
     6154.140   &    3623.26    &  1.04   \\
     6164.119   &    671.85      &  0.94   \\
     6173.095   &   -2506.32    &  1.18  \\
\hline
\hline
\end{tabular}}
\caption{Keck HIRES RV measurements.}
\label{tab:keck_hires}
\end{table}

The \citet{johnson_06} survey measurements were obtained with the Hamilton echelle spectrometer \citep{vogt_87} at Lick Observatory using the 3m Shane Telescope and 0.6m Coud\'{e} Auxiliary Telescope (CAT). The first observations of HD~8375 began on Sept. 8, 2004 (Table 2). Subsequent to the identification of a long-term trend, we also began Doppler monitoring with the HIgh Resolution Echelle Spectrometer (HIRES; \citealt{vogt_94}) at Keck (Table 3). Fig. 1 shows the relative RV measurements used for this study, which span 8.0 years and include the \citet{bowler_10} data points along with seven recent measurements from HIRES. 

The additional RV measurements from HIRES allow us to refine the orbital solution of HD~8375~AB. Following the methodology of \citet{wright_howard_09}, we use the {\tt RVLIN} software package\footnote{{\tt RVLIN} and {\tt boottran} are available at http://exoplanets.org/code} to simultaneously fit the Lick and Keck velocities. Parameter uncertainties are derived following a ``boot-strap analysis" methodology described in Wang et al. 2012 with the {\tt boottran} software package. For this fit we assume no jitter and allow for an offset between Lick and Keck velocities. Parameters from our updated orbital solution are shown in Table 2.

The r.m.s. to our companion plus trend fit is 25 m/s, considerably higher than internal errors.  There is no indication that this star should exhibit such high levels of jitter, nor does a periodogram analysis and manual search for a second, shorter period companion reveal any significant signal in the residuals. The residual scatter could be due to a small amount of flux from HD 8735 B contaminating our spectra and complicating the forward modeling procedure in our precise Doppler analysis. However, visual inspection of stellar template spectra does not reveal any anomalous features. HD 8375~A appears to dominate the spectrum of the entire system. Using a cross-correlation analysis, we place an upper-limit on the amount of flux contamination in the $\lambda=0.50-0.62 \mu$m wavelength range at the 1\% level. 

\begin{figure*}[!t]
\begin{center}
\includegraphics[height=3.3in]{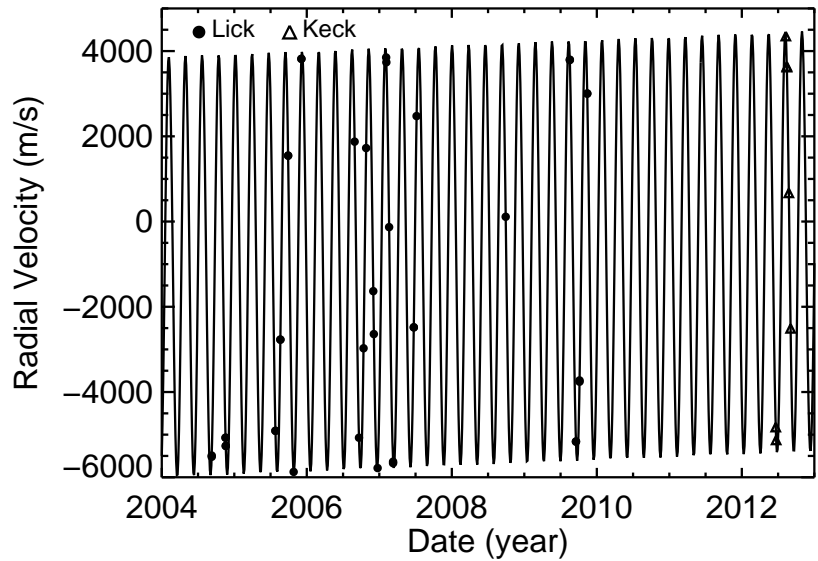} 
\caption{Precise Doppler measurements of HD~8375. An SB1 stellar pair with 84 day period exhibits a subtle long-term acceleration of $67.4\pm2.2$ m/s/yr suggesting the presence of a distant tertiary companion.} 
\end{center}\label{fig:rvs}
\end{figure*} 

\begin{figure*}[!t]
\begin{center}
\includegraphics[height=3.3in]{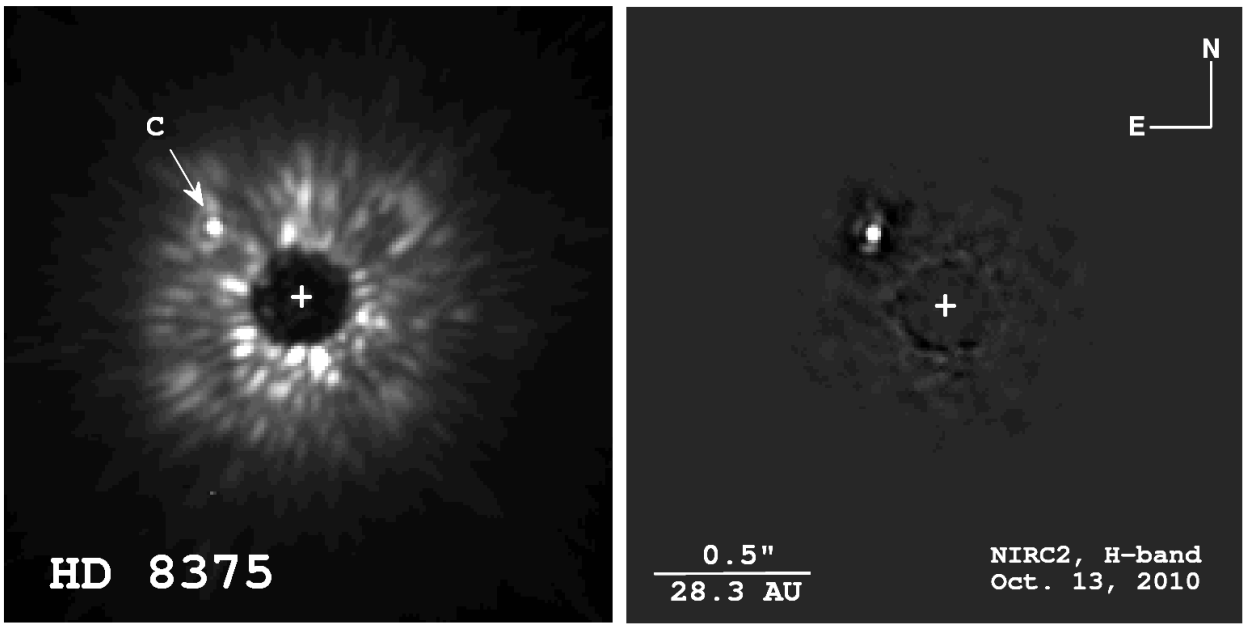} 
\caption{Discovery images of HD~8375~C taken in H-band at Keck with NIRC2 on Oct. 13, 2010 UT. The companion, discernible in individual frames (left), is unambiguously recovered following PSF subtraction (right). A cross denotes the location of the primary star behind the coronagraph.} 
\end{center}\label{fig:image}
\end{figure*} 

There is a hint of curvature in the Lick velocities (which the Keck velocities cannot constrain because of the offset between the telescopes and short time baseline of that particular data set). The velocities increase by $\approx$40 m/s from 2007 to 2010, a jump that would normally be highly significant, but may be spurious in this case given the large, unexplained r.m.s. scatter about the companion plus trend fit. Continued Doppler monitoring is required to assess the change in slope of the observed acceleration over the next several years.

\begin{table}[!t]
\centerline{
\begin{tabular}{lc}
\hline
\hline
\multicolumn{2}{c}{HD~8375~A}      \\
\hline
\hline
Mass [$M_{\odot}$]           &    $1.45\pm0.12$       \\
Radius [$R_{\odot}$]         &     $3.7\pm0.2$           \\
Luminosity [$L_{\odot}$]    &   $8.5\pm0.5$             \\
$\mbox{[Fe/H]}$                 &      $-0.13\pm0.04$        \\
log g [cm $\mbox{s}^{-2}$]    &  $3.46\pm0.06$       \\
$T_{\rm eff}$ [$K$]             &    $5103\pm44$                  \\
Spectral Type                     &      G8IV                                \\
v sini   [km/s]                        &   $2.0\pm0.5$                                             \\
\hline
\hline
\multicolumn{2}{c}{HD~8375~B Orbit}   \\
\hline
\hline        
$m_B\sin(i)$ [$M_{\odot}$]   &  $0.137\pm0.007$   \\
$P$ [day]      &   $83.9428\pm0.0013$  \\
$K$ [m/s]      &   $4931.0\pm3.0$  \\
$e$                &   $0.0179\pm0.0008$  \\
$\omega$ [$^{\circ}$]  &  $329.50\pm2.13$ \\
$t_p$ [JD-2,450,000]  &  $4125.01\pm0.50$   \\
$\gamma$ [m/s]          & $-960.4\pm14.0$  \\     
$\mbox{dv/dt}$ [m/s/yr]  &   $67.4\pm2.2$   \\
\hline
\hline
\end{tabular}}
\caption{(top) Physical properties of HD~8375~A derived from SME and theoretical isochrones using HIRES template spectra \citep{valenti_fischer_05}. We notice no contamination in the $\lambda=0.50-0.62 \mu$m range from HD~8375~BC. (bottom) Refined orbital parameters for HD~8375~AB using recent Doppler measurements. Model variables include the period ($P$), RV semi-amplitude ($K$), eccentricity ($e$), argument of periastron ($\omega$), time of periastron passage ($t_p$), RV instrument offset from Lick to Keck ($\gamma$), and Doppler acceleration ($\mbox{dv/dt}$). We calculate the minimum mass of HD~8375~B from RV measurements using the $1.45\pm0.12M_{\odot}$ estimated mass of HD~8375~A.}
\label{tab:hd8375}
\end{table}

\subsection{Primary Star Properties} 
Since close inspection of HIRES observations reveals virtually no contamination from HD~8375~BC, we are justified in deriving bulk physical properties of the primary star. Stellar (template) spectra, taken with the iodine gas cell removed from the optical path, were analyzed using the LTE spectral synthesis code {\it Spectroscopy Made Easy} (SME) described in \citet{valenti_fischer_05}. SME provides an estimate of the stellar effective temperature ($T_{\rm eff}$), surface gravity ($\log \rm{g}$), metallicity ($\mbox{[Fe/H]}$), and projected rotational velocity ($v \sin i$). Table 4 lists the spectral-type and physical properties of the primary star derived from spectral fitting along with comparison to theoretical isochrones. The SME model yields an excellent fit with a reduced $\chi^2$ close to one, independently verifying that very little flux from the secondary and tertiary are contaminating the spectra at visible wavelengths. 

\subsection{High-Contrast Imaging}\label{sec:imaging}
Knowing that HD~8375 must have a distant body orbiting the central binary pair, we acquired high-contrast images of the system with NIRC2 (instrument PI: Keith Matthews) using the Keck II AO system \citep{wizinowich_00} on Oct. 13, 2010 UT. First epoch images were taken using the H-band filter. The star (spatially unresolved binary) was placed behind the 300 mas diameter coronagraph spot. We used the angular differential imaging (ADI) technique to discriminate between residual scattered starlight and companions \citep{marois_06}. 

Fig. 2 shows images taken before and after speckle suppression. The companion has a brightness comparable to the static speckle pattern though can be seen in individual pre-processed frames. Using the locally optimized combination of images (LOCI) algorithm \citep{lafreniere_07}, and derotating frames to align and stack light from off-axis sources, we are able to identify the candidate located within close vicinity and to the north-east of the primary star(s). Follow-up observations were taken on Aug. 31, 2011 UT and August 25, 2012 UT with complementary filters to obtain color information and determine whether the candidate tertiary is associated with the primary. We note that the {\it a priori} likelihood that the point source is a false-positive is low, given the small (326 mas) angular separation and location of HD~8375 relative to the galactic plane. 

\section{ASTROMETRY}
Our astrometric observations consist of three epochs taken over an 1.9 year time baseline (Table~5). The proper-motion of HD~8375 is high (see Table 1), allowing us to easily determine whether the companion shares the same space motion as the primary. We measured an accurate angular separation and position angle using the technique described in \citep{crepp_12a}. We first fit Gaussian functions to the stellar and companion point-spread functions to locate their centroids in each frame. We then correct for distortion in the NIRC2 focal plane using publicly available solutions provided by Keck Observatory's astrometry support page \citep{ghez_08}.\footnote{http://www2.keck.hawaii.edu/inst/nirc2/forReDoc/post$\_$observing/dewarp/} The results are averaged and uncertainty in the separation and position angle is taken as the standard deviation, taking into account uncertainty in the plate scale and orientation of the array by propagating these errors to the final calculated position. 

\begin{table*}[t]
\centerline{
\begin{tabular}{lcccc}
\hline
\hline
Date [UT]    &   JD-2,450,000         &      $\rho$ [mas]      &    P. A. [$^{\circ}$]    &     Proj. Sep. [AU]    \\
\hline
\hline        
Oct. 13, 2010      &   5,482.94          &   $326.0\pm4.5$     &  $48.9\pm0.8$    &   $18.5\pm0.5$     \\
Aug. 31, 2011      &  5,804.06         &    $309.5\pm2.8$     &   $45.3\pm0.4$    &   $17.5\pm0.5$    \\
Aug. 25, 2012     &   6,165.10         &    $285.5\pm1.1$     &   $39.9\pm0.3$    &   $16.2\pm0.4$    \\
\hline
\hline
\end{tabular}}
\caption{Summary of astrometric measurements. HD~8375~C is moving in a clock-wise direction relative to HD~8375~AB and its projected separation is decreasing rapidly with time.}
\label{tab:astrometry}
\end{table*}

\begin{figure}[!t]
\begin{center}
\includegraphics[height=2.5in]{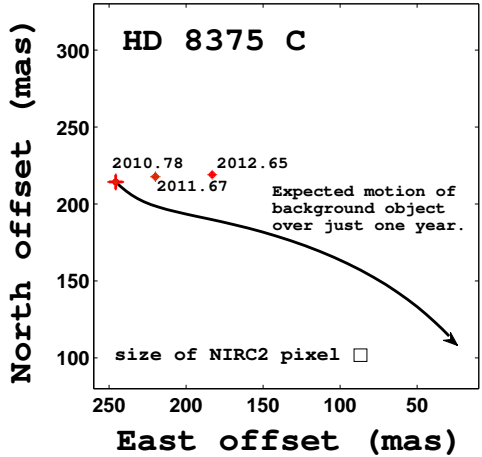} 
\caption{Astrometric measurements (red-crosses) demonstrating that HD~8375~C is co-moving with HD~8375~AB. Axes correspond to the measured angular separation (offset) of HD~8375~C from HD~8375~AB. Solid-curves show the path that a distant background object with zero proper-motion would follow from October 13, 2010 through August 31, 2011 accounting for stellar proper-motion and parallactic motion. Our measurements span a time frame twice as long, from October 13, 2010 through August 25, 2012; we have zoomed in to show the size of astrometric uncertainties.} 
\end{center}\label{fig:astrometry}
\end{figure} 

Fig.~3 shows multi-epoch astrometry measurements plotted against the expected motion of a distant background object. We find that HD~8375~C is clearly associated with HD~8375~AB. The tertiary has a projected separation of $16.2\pm0.4$ AU as of August, 2012, and appears to exhibit orbital motion in a clock-wise direction (north up, east left) as indicated by a slow systematic change in the separation and position angle. 


\section{COMPANION MASS}
We estimate the companion mass by comparing its brightness to: (i) late-type dwarfs using \citet{dotter_08} theoretical evolutionary tracks (the Dartmouth models), (ii) empirical relations from \cite{delfosse_00} that correlate absolute magnitudes with dynamical masses, and (iii) our lower-limit constraint on the mass from orbital dynamics. Differential magnitudes, apparent magnitudes, and absolute magnitudes are listed in Table~6. 

We find that HD~8375~C has a mass of $0.522\pm0.039M_{\odot}$ and $0.549\pm0.012M_{\odot}$ based on $H$-band and $K_s$-band photometry respectively \citep{dotter_08}. These values take into consideration the metallicity of HD~8375~A (assuming each star has the same chemical composition), and are consistent to within $1\sigma$. Using Table~5 from \citet{kraus_hillenbrand_07}, HD~8375~C has colors and brightness most consistent with an M1-dwarf. 

Empirical mass-luminosity relations have a relatively small scatter for M-dwarf stars, often yielding mass predictions accurate to 10\%. Substituting absolute magnitudes from Table~6 into the polynomial-fit relations derived by \citealt{delfosse_00}, we find that HD~8375~C has a mass of $0.44\pm0.11M_{\odot}$ and $0.58\pm0.04M_{\odot}$ for the $H$ and $K_s$ filters respectively. These values are also consistent with one another and the \cite{dotter_08} theoretical evolutionary models at the $1\sigma$ level. We find that this consistency is only true when it is assumed the tertiary shares the same metallicity as the primary. Invoking a solar metallicity for instance introduces an $>1\sigma$ discrepancy between the mass obtained from empirical relations versus atmospheric models. One can interpret this result as suggestive of a common formation origin, however the evidence is indirect.  

We can place a lower-limit on the companion dynamical mass by combining the RV trend with our measurements of HD~8375~C's projected separation \citep{torres_99,liu_02}. Using our updated Doppler acceleration, $dv/dt=67.4\pm2.2$ m$\:$s$^{-1}$yr$^{-1}$, and first epoch direct astrometry from October 2010, we find that HD~8375~C has a minimum dynamical mass of $m_{\rm dyn}=0.319\pm0.022M_{\odot}$. This value accounts for uncertainty in the stellar parallax, measured angular separation, and RV acceleration. We thus adopt a minimum mass of $m_{\rm dyn}=0.297M_{\odot}$. This lower-limit is consistent with masses derived above from photometry and empirical inference (Table 6). Subsequent observations will further constrain the mass (placing an upper-limit as well) when the HD~8375~C astrometry shows significant curvature. 

We can also perform a complementary analysis and self-consistency check to estimate the true (physical) separation of HD~8375~C from (the center of mass of) HD~8375~AB (c.f., \citealt{howard_10}). If HD~8375~C has a mass of $m_{\rm model}=0.547\pm0.011M_{\odot}$, which is the weighted average of our $H$ and $K$ estimates from photometry, then the instantaneous orbital separation of HD~8375~C (August 2012) is $16.5\pm0.5$ AU, only marginally larger than the $16.2\pm0.4$ AU projected separation. Note that comparing these values does not yield the orbit inclination. 

\begin{table}[!t]
\centerline{
\begin{tabular}{lc}
\hline
\hline
                         \multicolumn{2}{c}{HD~8375~C}   \\
\hline
\hline        
$\Delta J$                          &      $>4.85$                    \\
$\Delta H$                         &      $5.35\pm0.15$       \\
$\Delta K_s$                     &      $4.88\pm0.05$       \\
$J$                                         &     $> 9.67$                \\
$H$                                        &     $9.56\pm0.28$     \\
$K_s$                                    &    $9.17\pm0.06$     \\
$M_J$                                    &    $>5.90$                  \\
$M_H$                                   &    $5.78\pm0.29$     \\
$M_{K_s}$                            &     $5.40\pm0.08$    \\
Spec. Type                            &      $\approx$M1V                      \\
$m_{\rm dyn}$ [$M_{\odot}$]   &    $>0.297$                  \\
$m_{\rm model}$ [$M_{\odot}$]        &    $0.522\pm0.039$ ($H$-band) \\ 
                                                          &    $0.549\pm0.012$ ($K_s$-band)   \\
$m_{\rm empirical}$ [$M_{\odot}$]    &    $0.44\pm0.11$ ($H$-band) \\
                                                          &    $0.58\pm0.04$ ($K_s$-band)  \\
\hline
\end{tabular}}
\caption{Tertiary companion magnitude difference, apparent magnitude, absolute magnitude, estimated spectral-type, mass constraint from dynamics ($m_{dyn}$), and estimated mass from: (i) photometry using the \citet{dotter_08} theoretical atmospheric models ($m_{model}$), and (ii) empirical $M_H$-mass and $M_K$-mass relations from \citet{delfosse_00} ($m_{\rm empirical}$). Measurements are made relative to the combined (spatially unresolved) light from HD~8375~AB. The tertiary spectral-type is estimated based on its $H-K_s$ color.}
\label{tab:compprops}
\end{table}

\section{SUMMARY}
We present the second discovery of the TRENDS high-contrast imaging program. Using multi-epoch NIRC2 AO observations at Keck, we have directly imaged the companion responsible for accelerating HD~8375~AB, a single-line spectroscopic binary with an 83.9 day period. The tertiary, HD~8375~C, is only slightly brighter than residual scattered light from the primary. Our multi-band observations indicate that HD~8375~C has brightness and colors consistent with an $\approx$M1 dwarf. Combining imaging observations with precise Doppler measurements, we derive a firm lower-limit of $0.297M_{\odot}$ for the tertiary. HD~8375~C shows measurable orbital motion in a clockwise direction over a 1.9 year time baseline.

We find that the estimated mass of HD~8375~C is consistent between evolutionary models and empirical mass-luminosity relations, but only once the metallicity of HD~8375~A is taken into account (assuming a common chemical composition). A dynamical mass with fractional error $<10\%$ is possible with continued follow-up observations, and will be sufficient to identify any small systematic errors in M-dwarf theoretical atmospheric models. 

Follow-up moderate resolution spectroscopy using an integral-field unit will help verify the spectral-type we have assigned to HD~8375~C. Further, near-infrared spectra of the primary can place strong constraints on the properties of HD~8375~B, which is difficult to detect at visible wavelengths, via a combined light analysis. We have recently acquired low-resolution JH spectra of HD~8375~C using Project 1640 at Palomar \citep{hinkley_11_PASP,crepp_11,pueyo_12}. These observations and analysis will be presented in a separate paper.  

HD~8375 is an exemplar hierarchical triple-star system that was initially characterized as a spectroscopic binary. Compared to single stars and binaries, joint Doppler and imaging observations can address more subtle issues involving the star-formation process, such as angular momentum orientation, mass and separation ratios, and the Kozai mechanism. Triple stars represent approximately $8\%$ of all stellar systems \citep{raghavan_10}. Thus, we anticipate that the TRENDS high-contrast survey will uncover additional interesting triple-systems for which to study in detail.

\section{ACKNOWLEDGEMENTS}
The TRENDS high-contrast science program is supported in part by NASA Origins grant NNX13AB03G. The data presented herein were obtained at the W.M. Keck Observatory, which is operated as a scientific partnership among the California Institute of Technology, the University of California and the National Aeronautics and Space Administration. The Observatory was made possible by the generous financial support of the W.M. Keck Foundation. The Center for Exoplanets and Habitable Worlds is supported by the Pennsylvania State University, the Eberly College of Science, and the Pennsylvania Space Grant Consortium.

%

\begin{small}
\bibliographystyle{jtb}
\bibliography{TRENDSII_v3.bib}

\begin{thebibliography}{}

\bibitem[\protect\astroncite{{Apps} et~al.}{2010}]{apps_10}
{Apps}, K., {Clubb}, K.~I., {Fischer}, D.~A., {Gaidos}, E., {Howard}, A.,
  {Johnson}, J.~A., {Marcy}, G.~W., {Isaacson}, H., {Giguere}, M.~J.,
  {Valenti}, J.~A., {Rodriguez}, V., {Chubak}, C., and {Lepine}, S. (2010) ,
\newblock {\em \pasp} {\bf 122}, 156

\bibitem[\protect\astroncite{{Bate}}{2009}]{bate_09}
{Bate}, M.~R. (2009) ,
\newblock {\em \mnras} {\bf 392}, 590

\bibitem[\protect\astroncite{{Beavers} and {Eitter}}{1986}]{beavers_eitter_86}
{Beavers}, W.~I. and {Eitter}, J.~J. (1986) ,
\newblock {\em \apjs} {\bf 62}, 147

\bibitem[\protect\astroncite{{Boden} et~al.}{2006}]{boden_06}
{Boden}, A.~F., {Torres}, G., and {Latham}, D.~W. (2006) ,
\newblock {\em \apj} {\bf 644}, 1193

\bibitem[\protect\astroncite{{Bowler} et~al.}{2010}]{bowler_10}
{Bowler}, B.~P., {Johnson}, J.~A., {Marcy}, G.~W., {Henry}, G.~W., {Peek},
  K.~M.~G., {Fischer}, D.~A., {Clubb}, K.~I., {Liu}, M.~C., {Reffert}, S.,
  {Schwab}, C., and {Lowe}, T.~B. (2010) ,
\newblock {\em \apj} {\bf 709}, 396

\bibitem[\protect\astroncite{{Crepp} et~al.}{2010}]{crepp_10}
{Crepp}, J., {Serabyn}, E., {Carson}, J., {Ge}, J., and {Kravchenko}, I. (2010)
  ,
\newblock {\em \apj} {\bf 715}, 1533

\bibitem[\protect\astroncite{{Crepp} et~al.}{2012a}]{crepp_12a}
{Crepp}, J.~R., {Johnson}, J.~A., {Fischer}, D.~A., {Howard}, A.~W., {Marcy},
  G.~W., {Wright}, J.~T., {Isaacson}, H., {Boyajian}, T., {von Braun}, K.,
  {Hillenbrand}, L.~A., {Hinkley}, S., {Carpenter}, J.~M., and {Brewer}, J.~M.
  (2012a) ,
\newblock {\em \apj} {\bf 751}, 97

\bibitem[\protect\astroncite{{Crepp} et~al.}{2012b}]{crepp_12b}
{Crepp}, J.~R., {Johnson}, J.~A., {Howard}, A.~W., {Marcy}, G.~W., {Fischer},
  D.~A., {Hillenbrand}, L.~A., {Yantek}, S.~M., {Delaney}, C.~R., {Wright},
  J.~T., {Isaacson}, H.~T., and {Montet}, B.~T. (2012b) ,
\newblock {\em \apj} {\bf 761}, 39

\bibitem[\protect\astroncite{{Crepp} et~al.}{2011}]{crepp_11}
{Crepp}, J.~R., {Pueyo}, L., {Brenner}, D., {Oppenheimer}, B.~R., {Zimmerman},
  N., {Hinkley}, S., {Parry}, I., {King}, D., {Vasisht}, G., {Beichman}, C.,
  {Hillenbrand}, L., {Dekany}, R., {Shao}, M., {Burruss}, R., {Roberts}, L.~C.,
  {Bouchez}, A., {Roberts}, J., and {Soummer}, R. (2011) ,
\newblock {\em \apj} {\bf 729}, 132

\bibitem[\protect\astroncite{{de Medeiros} and {Mayor}}{1999}]{deMedeiros_99}
{de Medeiros}, J.~R. and {Mayor}, M. (1999) ,
\newblock {\em \aaps} {\bf 139}, 433

\bibitem[\protect\astroncite{{Delfosse} et~al.}{2000}]{delfosse_00}
{Delfosse}, X., {Forveille}, T., {S{\'e}gransan}, D., {Beuzit}, J.-L., {Udry},
  S., {Perrier}, C., and {Mayor}, M. (2000) ,
\newblock {\em \aap} {\bf 364}, 217

\bibitem[\protect\astroncite{{Dotter} et~al.}{2008}]{dotter_08}
{Dotter}, A., {Chaboyer}, B., {Jevremovi{\'c}}, D., {Kostov}, V., {Baron}, E.,
  and {Ferguson}, J.~W. (2008) ,
\newblock {\em \apjs} {\bf 178}, 89

\bibitem[\protect\astroncite{{Duch{\^e}ne} et~al.}{2006}]{duchene_06}
{Duch{\^e}ne}, G., {Beust}, H., {Adjali}, F., {Konopacky}, Q.~M., and {Ghez},
  A.~M. (2006) ,
\newblock {\em \aap} {\bf 457}, L9

\bibitem[\protect\astroncite{{Dupuy} et~al.}{2010}]{dupuy_10}
{Dupuy}, T.~J., {Liu}, M.~C., {Bowler}, B.~P., {Cushing}, M.~C., {Helling}, C.,
  {Witte}, S., and {Hauschildt}, P. (2010) ,
\newblock {\em \apj} {\bf 721}, 1725

\bibitem[\protect\astroncite{{Ghez} et~al.}{2008}]{ghez_08}
{Ghez}, A.~M., {Salim}, S., {Weinberg}, N.~N., {Lu}, J.~R., {Do}, T., {Dunn},
  J.~K., {Matthews}, K., {Morris}, M.~R., {Yelda}, S., {Becklin}, E.~E.,
  {Kremenek}, T., {Milosavljevic}, M., and {Naiman}, J. (2008) ,
\newblock {\em \apj} {\bf 689}, 1044

\bibitem[\protect\astroncite{{Hinkley} et~al.}{2011}]{hinkley_11_PASP}
{Hinkley}, S., {Oppenheimer}, B.~R., {Zimmerman}, N., {Brenner}, D., {Parry},
  I.~R., {Crepp}, J.~R., {Vasisht}, G., {Ligon}, E., {King}, D., {Soummer}, R.,
  {Sivaramakrishnan}, A., {Beichman}, C., {Shao}, M., {Roberts}, L.~C.,
  {Bouchez}, A., {Dekany}, R., {Pueyo}, L., {Roberts}, J.~E., {Lockhart}, T.,
  {Zhai}, C., {Shelton}, C., and {Burruss}, R. (2011) ,
\newblock {\em \pasp} {\bf 123}, 74

\bibitem[\protect\astroncite{{Howard} et~al.}{2010}]{howard_10}
{Howard}, A.~W., {Johnson}, J.~A., {Marcy}, G.~W., {Fischer}, D.~A., {Wright},
  J.~T., {Bernat}, D., {Henry}, G.~W., {Peek}, K.~M.~G., {Isaacson}, H.,
  {Apps}, K., {Endl}, M., {Cochran}, W.~D., {Valenti}, J.~A., {Anderson}, J.,
  and {Piskunov}, N.~E. (2010) ,
\newblock {\em \apj} {\bf 721}, 1467

\bibitem[\protect\astroncite{{Johnson} et~al.}{2011}]{johnson_11}
{Johnson}, J.~A., {Clanton}, C., {Howard}, A.~W., {Bowler}, B.~P., {Henry},
  G.~W., {Marcy}, G.~W., {Crepp}, J.~R., {Endl}, M., {Cochran}, W.~D.,
  {MacQueen}, P.~J., {Wright}, J.~T., and {Isaacson}, H. (2011) ,
\newblock {\em \apjs} {\bf 197}, 26

\bibitem[\protect\astroncite{{Johnson} et~al.}{2006}]{johnson_06}
{Johnson}, J.~A., {Marcy}, G.~W., {Fischer}, D.~A., {Henry}, G.~W., {Wright},
  J.~T., {Isaacson}, H., and {McCarthy}, C. (2006) ,
\newblock {\em \apj} {\bf 652}, 1724

\bibitem[\protect\astroncite{{Konopacky} et~al.}{2010}]{konopacky_10}
{Konopacky}, Q.~M., {Ghez}, A.~M., {Barman}, T.~S., {Rice}, E.~L., {Bailey},
  III, J.~I., {White}, R.~J., {McLean}, I.~S., and {Duch{\^e}ne}, G. (2010) ,
\newblock {\em \apj} {\bf 711}, 1087

\bibitem[\protect\astroncite{{Kraus} and
  {Hillenbrand}}{2007}]{kraus_hillenbrand_07}
{Kraus}, A.~L. and {Hillenbrand}, L.~A. (2007) ,
\newblock {\em \aj} {\bf 134}, 2340

\bibitem[\protect\astroncite{{Lafreni{\`e}re} et~al.}{2007}]{lafreniere_07}
{Lafreni{\`e}re}, D., {Doyon}, R., {Marois}, C., {Nadeau}, D., {Oppenheimer},
  B.~R., {Roche}, P.~F., {Rigaut}, F., {Graham}, J.~R., {Jayawardhana}, R.,
  {Johnstone}, D., {Kalas}, P.~G., {Macintosh}, B., and {Racine}, R. (2007) ,
\newblock {\em \apj} {\bf 670}, 1367

\bibitem[\protect\astroncite{{Liu} et~al.}{2002}]{liu_02}
{Liu}, M.~C., {Fischer}, D.~A., {Graham}, J.~R., {Lloyd}, J.~P., {Marcy},
  G.~W., and {Butler}, R.~P. (2002) ,
\newblock {\em \apj} {\bf 571}, 519

\bibitem[\protect\astroncite{{Marois} et~al.}{2006}]{marois_06}
{Marois}, C., {Lafreni{\`e}re}, D., {Doyon}, R., {Macintosh}, B., and {Nadeau},
  D. (2006) ,
\newblock {\em \apj} {\bf 641}, 556

\bibitem[\protect\astroncite{{Pueyo} et~al.}{2012}]{pueyo_12}
{Pueyo}, L., {Crepp}, J.~R., {Vasisht}, G., {Brenner}, D., {Oppenheimer},
  B.~R., {Zimmerman}, N., {Hinkley}, S., {Parry}, I., {Beichman}, C.,
  {Hillenbrand}, L., {Roberts}, L.~C., {Dekany}, R., {Shao}, M., {Burruss}, R.,
  {Bouchez}, A., {Roberts}, J., and {Soummer}, R. (2012) ,
\newblock {\em \apjs} {\bf 199}, 6

\bibitem[\protect\astroncite{{Raghavan} et~al.}{2010}]{raghavan_10}
{Raghavan}, D., {McAlister}, H.~A., {Henry}, T.~J., {Latham}, D.~W., {Marcy},
  G.~W., {Mason}, B.~D., {Gies}, D.~R., {White}, R.~J., and {ten Brummelaar},
  T.~A. (2010) ,
\newblock {\em ArXiv e-prints}

\bibitem[\protect\astroncite{{Schaefer} et~al.}{2012}]{schaefer_12}
{Schaefer}, G.~H., {Prato}, L., {Simon}, M., and {Zavala}, R.~T. (2012) ,
\newblock {\em \apj} {\bf 756}, 120

\bibitem[\protect\astroncite{{Skrutskie} et~al.}{2006}]{skrutskie_06}
{Skrutskie}, M.~F., {Cutri}, R.~M., {Stiening}, R., {Weinberg}, M.~D.,
  {Schneider}, S., {Carpenter}, J.~M., {Beichman}, C., {Capps}, R., {Chester},
  T., {Elias}, J., {Huchra}, J., {Liebert}, J., {Lonsdale}, C., {Monet}, D.~G.,
  {Price}, S., {Seitzer}, P., {Jarrett}, T., {Kirkpatrick}, J.~D., {Gizis},
  J.~E., {Howard}, E., {Evans}, T., {Fowler}, J., {Fullmer}, L., {Hurt}, R.,
  {Light}, R., {Kopan}, E.~L., {Marsh}, K.~A., {McCallon}, H.~L., {Tam}, R.,
  {Van Dyk}, S., and {Wheelock}, S. (2006) ,
\newblock {\em \aj} {\bf 131}, 1163

\bibitem[\protect\astroncite{{Snowden} and {Young}}{2005}]{snowden_05}
{Snowden}, M.~S. and {Young}, A. (2005) ,
\newblock {\em \apjs} {\bf 157}, 126

\bibitem[\protect\astroncite{{Tokovinin}}{2004}]{tokovinin_04}
{Tokovinin}, A. (2004) ,
\newblock In {\em Revista Mexicana de Astronomia y Astrofisica Conference
  Series}.  (C. {Allen} and C. {Scarfe} eds.), Vol.~21 of {\em Revista Mexicana
  de Astronomia y Astrofisica Conference Series}, pp. 7--14

\bibitem[\protect\astroncite{{Torres}}{1999}]{torres_99}
{Torres}, G. (1999) ,
\newblock {\em \pasp} {\bf 111}, 169

\bibitem[\protect\astroncite{{Valenti} and
  {Fischer}}{2005}]{valenti_fischer_05}
{Valenti}, J.~A. and {Fischer}, D.~A. (2005) ,
\newblock {\em \apjs} {\bf 159}, 141

\bibitem[\protect\astroncite{{van Leeuwen}}{2007}]{van_leeuwen_07}
{van Leeuwen}, F. (2007) ,
\newblock {\em \aap} {\bf 474}, 653

\bibitem[\protect\astroncite{{Vogt}}{1987}]{vogt_87}
{Vogt}, S.~S. (1987) ,
\newblock {\em \pasp} {\bf 99}, 1214

\bibitem[\protect\astroncite{{Vogt} et~al.}{1994}]{vogt_94}
{Vogt}, S.~S., {Allen}, S.~L., {Bigelow}, B.~C., {Bresee}, L., {Brown}, B.,
  {Cantrall}, T., {Conrad}, A., {Couture}, M., {Delaney}, C., {Epps}, H.~W.,
  {Hilyard}, D., {Hilyard}, D.~F., {Horn}, E., {Jern}, N., {Kanto}, D.,
  {Keane}, M.~J., {Kibrick}, R.~I., {Lewis}, J.~W., {Osborne}, J.,
  {Pardeilhan}, G.~H., {Pfister}, T., {Ricketts}, T., {Robinson}, L.~B.,
  {Stover}, R.~J., {Tucker}, D., {Ward}, J., and {Wei}, M.~Z. (1994) ,
\newblock In {\em Society of Photo-Optical Instrumentation Engineers (SPIE)
  Conference Series}.  ({D.~L.~Crawford \& E.~R.~Craine} ed.), Vol. 2198 of
  {\em Society of Photo-Optical Instrumentation Engineers (SPIE) Conference
  Series}, pp. 362--+

\bibitem[\protect\astroncite{{Wizinowich} et~al.}{2000}]{wizinowich_00}
{Wizinowich}, P., {Acton}, D.~S., {Shelton}, C., {Stomski}, P., {Gathright},
  J., {Ho}, K., {Lupton}, W., {Tsubota}, K., {Lai}, O., {Max}, C., {Brase}, J.,
  {An}, J., {Avicola}, K., {Olivier}, S., {Gavel}, D., {Macintosh}, B., {Ghez},
  A., and {Larkin}, J. (2000) ,
\newblock {\em \pasp} {\bf 112}, 315

\bibitem[\protect\astroncite{{Wright} and {Howard}}{2009}]{wright_howard_09}
{Wright}, J.~T. and {Howard}, A.~W. (2009) ,
\newblock {\em \apjs} {\bf 182}, 205

\bibitem[\protect\astroncite{{Wright} et~al.}{2009}]{wright_09}
{Wright}, J.~T., {Upadhyay}, S., {Marcy}, G.~W., {Fischer}, D.~A., {Ford},
  E.~B., and {Johnson}, J.~A. (2009) ,
\newblock {\em \apj} {\bf 693}, 1084

\end{thebibliography}
\end{small}

\end{document}